\documentclass[cits]{PoS}

\title{Molecular tori in AGN:\\ a search using excited states of OH}

\ShortTitle{AGN Tori}

\author{\speaker{C. M. Violette Impellizzeri}\\
        Max-Planck-Institut f\"ur Radiostronomie, Bonn, Germany \\
        E-mail: \email{violette@mpifr-bonn.mpg.de}}

\author{Alan L. Roy, Christian Henkel \\
        Max-Planck-Institut f\"ur Radiostronomie, Bonn, Germany \\
        E-mail: \email{aroy@mpifr-bonn.mpg.de}, \email{chenkel@mpifr-bonn.mpg.de} }

\abstract{One of the fundamental concepts in the unified scheme of AGN is 
that both Seyfert~1 and Seyfert~2 galaxies harbour supermassive
nuclear engines blocked from direct view by an optically and
geometrically thick torus. If the pressure is sufficiently high, the
torus should mostly be molecular. Although molecular rings with
diameters of a few hundred parsecs are common, the expected small
scale tori ($<$10\,pc) have been difficult to detect. Searches for
absorption lines of common molecules like CO and OH have mostly
yielded non-detections. Before concluding that tori are not molecular,
radiative excitation effects, in which coupling to the nonthermal
continuum can suppress the opacity in the lowest transitions, deserve
some attention and influence our selection of the most favourable
transitions to observe. To explore these effects, we modified the
search strategy by looking for the higher excited rotational states of
OH and by selecting a sample of 31 Seyfert~2 galaxies which are known
to have a high X-ray absorbing column.  We present here the results of
single dish observations of the transitions at 6031\,MHz and
6035\,MHz, yielding detections in five sources. We also present a
spectral line VLBI observation carried out at 13.4\,GHz towards the
core of Cygnus~A, yielding a tentative detection. }

\FullConference{The 8th European VLBI Network Symposium on New Developments in VLBI Science and Technology
                and EVN Users Meeting  \\
		September 26-29 2006\\
		Torun, Poland}

\begin{document}
\section{Molecular Tori in AGN} 
Active galactic nuclei (AGN) come in two main types: those with and
those without broad optical line emission (type~1 and type~2 AGN,
respectively). In the unified scheme of active galactic nuclei, all
AGN are intrinsically similar; in the type 2 objects our view of the
central continuum source and the broad line region is blocked by a
significant column density of obscuring material \cite{1}. The
obscuring material is expected to be an approximately parsec-scale
torus of molecular gas, whose structure was predicted by e.g. Krolik
\& Begelman
\cite{3}. Despite many efforts to detect the expected molecular
absorption or emission in a number of surveys, only in very few cases
could molecular absorption be confirmed (e.g. Schmelz et al. 1986,
Baan et al. 1992, Staveley-Smith et al. 1992 \cite{4}, \cite{5},
\cite{6}). The unified scheme accounts for many observed properties, but
evidence for the existence of a molecular torus has been somewhat
indirect and the structure and extent of the obscuring material are
still poorly understood. As the nearest powerful FR II radio galaxy
($z$ = 0.0565), Cygnus~A has been the subject of a number of studies
that test the predictions of the unified scheme. Several lines of
evidence suggest that Cygnus~A harbours a quasar nucleus hidden from
direct view in the visible and soft X-rays by a dusty obscuring torus
\cite{7}. An X-ray absorbing column of $N_{\rm H}$ = (3.75 $\pm$ 0.73)
$\times$ 10$^{23}$\,cm$^{-2}$ is consistent with the notion of a
buried quasar in Cygnus~A \cite{8}.  Searches for 18\,cm OH and 6\,cm
H$_2$CO by Conway \& Blanco \cite{9} yielded non-detections at 1\%
optical depth \cite{9}. In 1994, Barvainis \& Antonucci failed to
detect CO J=0-1 and CO J=1-2 absorption, challenging the torus model
for Cygnus~A \cite{10}. The authors suggested three possible
solutions. First, it may be that there is no such torus. Second, the
sizes of the molecular clouds in the torus may be smaller than the
size of the background continuum source. Alternatively, the radio
continuum emission from the nucleus may radiatively excite the CO,
increasing the excitation temperature of the lower rotational levels
and suppressing the absorption optical depths in the lower transitions
\cite{2}.  Interestingly, Fuente et al. \cite{11} report
118\,GHz CO+ absorption, supporting CO excitation from radiative
excitation effects \cite{11}. Given the abundances of OH predicted in
some molecular torus models, the non-detection of OH absorption is
hard to explain unless this too is radiatively excited. Radiative
excitation effects have been predicted by Black (1997) also for the
18\,cm OH transitions \cite{12}, suggesting that the rotationally
excited doublets at 6\,GHz and 13.4\,GHz, so far neglected, could be
more profitable targets than the 1.6\,GHz lines.

\section{Observations} 

To explore the effects of radiative excitation we modified the search
strategy for molecular absorption, starting with a survey with the
100-metre telescope in Effelsberg to search for OH in the the
higher-order transitions at 6031\,MHz and 6035\,MHz in 31 type~2 AGN
(see Table~1) selected for having a known high X-ray absorbing column
($\geq$ $10^{22}$\,cm$^{-2}$) and for having $S_{\rm 6\,GHz}>$~50\,mJy
to ensure sufficient continuum core strength. We included some AGN
with polarized broad-line emission, which indicates an obscured broad
line region, and some additional sources with known OH absorption at
1.6\,GHz. The observations were carried out between August\,2003 and
February\,2004 with an average integration time of 3\,h per source and
40\,MHz bandwidth. Twenty of the stronger sources were followed up in
the 4.7\,GHz transitions with Effelsberg between June\,2004 and
August\,2004. \\ Four of our top candidates, Cygnus~A, Hydra~A,
NGC~1052 and NGC~1275, were too strong for single-dish spectroscopy,
demanding unachievably high spectral dynamic range and for two of
these we thus proposed an interferometric study. We looked for excited
OH at 13.4\,GHz in Cygnus~A and NGC~1052 with the VLBA, omitting
Hydra~A since its core was too weak at 13.4\,GHz. The observations
were made with two IFs of 16\,MHz bandwidth (corresponding to
357\,km/s per IF). The IFs were centred at 13.434\,GHz and
13.441\,GHz, respectively, with 256 channels per IF yielding a
spectral line resolution of 1.4\,km/s per channel. The data were
calibrated using standard phase and amplitude calibration, using
2005+403 and 2013+370 for bandpass calibration of Cygnus~A and 0423-01
and 3C84 for NGC~1052.

\begin{table}
{\scriptsize
\begin{tabular}{lcclcc|lcclcc}
Source     & Redshift & $S_{6\,cm}$ & $N_{\rm H}$ X-ray         &  Int. time & Det  & Source   & Redshift  & $S_{6\,cm}$ &  $N_{\rm H}$ X-ray     &  Int. time & Det    \\
           &   $z$    &     mJy     &     cm$^{-2}$             &   min      &      &          &           &             &                        &            &        \\ 
\hline
Hydra~A    &      0.055 &   154     &       -                   &   -        &      & NGC~5135 &   0.014   &  598        &    $> 10^{24}$          &  120       &  -    \\
Cygnus~A   &    0.056 &    1400     &     10$^{23.5}$           &   -        & -    & NGC~5506   &   0.006  &  160         &     3.4$\times 10^{23}$   &  230      &  - \\
NGC~1052   &    0.005 &    1500     &     $>$ 10$^{22}$         &   -        & -    & NGC~5793   &   0.011  &   96         &                           &  230      & abs \\
NGC~1068   &    0.004 &    1090     &     $>$10$^{25}$          &   -        & -    & NGC~6240   &   0.024  &   131        &   2.2$\times 10^{25}$     &  103      &  - \\
NGC~1167   &    0.002 &    243      &     $<$2$\times 10^{23}$  &   154      &  -   &  NGC~7130   &   0.016  &   62         &  $> 10^{24}$              &  180      &  - \\
NGC~1275   &    0.017 &    16623    &     1.5$\times 10^{23}$   &   -        &  -   & NGC~7674  .&  0.029   &   66.5       &   $> 10^{24}$             &   205     &  - \\
NGC~1365   &    0.006 &    191      &     2$\times 10^{24}$     &   180      &  -   &  Mrk~3      &   0.014  & 361          &   1.1$\times 10^{25}$     &  205      &  - \\
NGC~1808   &    0.003 &    207      &     3$\times 10^{23}$    &   180       &  -   & Mrk~231    &   0.042   & 414           &       -                    &  $\dag$  & abs  \\
NGC~2110   &    0.007 &    175      &     2.9$\times 10^{23}$  &   180       &  -   & Mrk~273    &   0.037   & 103           &       -                    &  $\dag$   & abs  \\
NGC~2639   &   0.040  &     54.5    &        -                 &   154       &  -   & Mrk~348    &  0.015   &  254         &   $10^{24}$                &  211     & - \\
NGC~2992   &   0.008  &    77       &    6.9$\times 10^{22}$   &   185       &  -   & Mrk~463    &  0.050   &  100          &  1.6$\times 10^{24}$      &  180     & - \\
NGC~3079   &   0.004  &    145      &    1.6$\times 10^{23}$   &    65      &  abs & Mrk~1210   &  0.014  &  45          &   $10^{25}$                &  180     & - \\
NGC~4151   &   0.003  &    125      &    2.2$\times 10^{22}$   &   255       & -    & Mrk~1073   &   0.024 &  44          &           -                 &   358    & - \\
NGC~4261   &   0.007  &    80       &        -                  &  180       &  abs  & F~01475-0740   & 0.177 &  127       &          -                  &  103    & -  \\    
NGC~4388   &   0.008  &    76       &    4.2$\times 10^{24}$    &  205       &  -    & IRAS~05414+5840  & 0.015 &  55      &          -                 &     180   & - \\
           &          &             &         -                  &            &       & IRAS~1345+1232   &  0.121 &   2160  &          -                &  180       & - \\
\end{tabular}
\caption{List of sources observed at 6\,GHz with the 100-m telescope in Effelsberg.  
	$\dag$ {\small sources observed by Henkel et al. (priv. comm.)} }
}
\end{table}

\begin{figure}[h!]
\centering
\includegraphics[width=150mm] {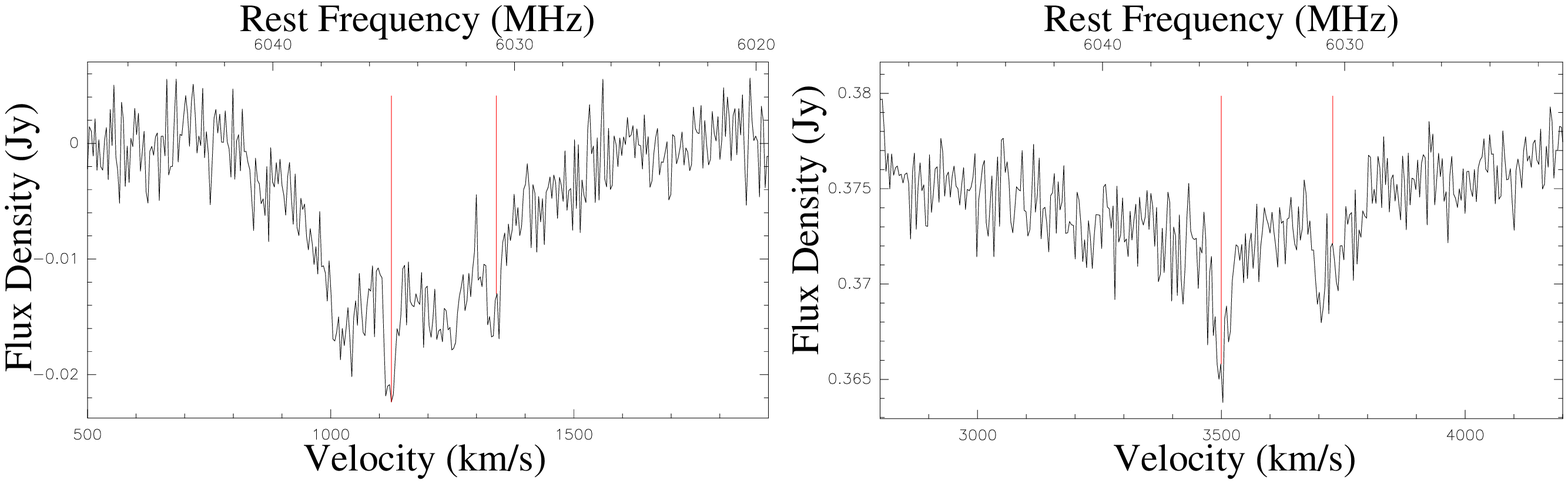}
\caption{Absorption spectra of NGC~3079 (left) and NGC~5793 (right). The red lines mark the position of the two main lines at 6031\,MHz and 6035\,MHz at the systemic velocity. The velocity scales refer to the 6035\,MHz line. }
\end{figure}	

\begin{figure}[h!]
\centering
\includegraphics[width=120mm] {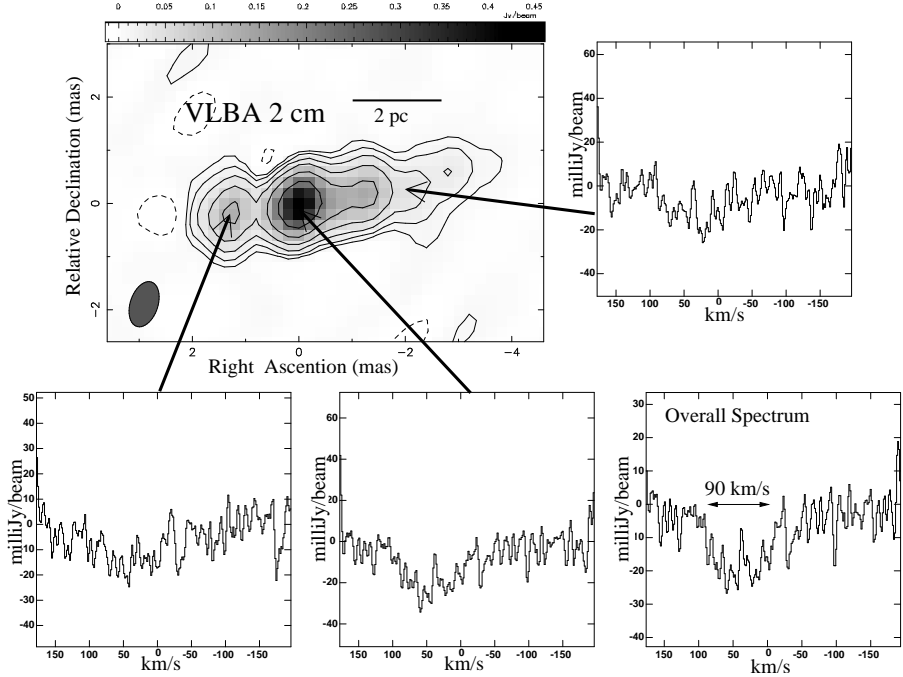}
\caption{ VLBA image of Cygnus~A at 2\,cm. The synthesized beam is (0.879$\times$0.52)\,mas$^2$. The image is contoured at -1.5, 1.5, 3, 6, 12, 24, 48, 96 percent of the map peak at 0.453\,Jy/beam. The panels show a montage with selected OH absorption spectra. The bottom right panel shows the absorption spectrum integrated over the overall source.  }
\end{figure}

\section{Results}
\subsection{An Effelsberg Survey}

Of the originally selected 31 Seyfert~2 galaxies listed in Table~1, 27
sources were observed at 6\,GHz and 20 sources were observed at
4.7\,GHz with the 100m telescope at Effelsberg. We detect absorption
at 6\,GHz towards five sources.  The spectra of two of them, NGC~3079
and NGC~5793, are shown in Fig.~1. Absorption is detected towards the
centre of the sources, with wide velocity components and narrow
troughs at the systemic velocity of the two lines. In NGC~3079 (Fig~1,
left panel), the absorption width is 800\,km/s, with two deeper
components corresponding to the two transitions (6031\,MHz and
6035\,MHz) at the systemic velocity. The maximum line opacity is
$\tau$\,=\,0.05. Two further lines are also observed corresponding to
the two transitions blueshifted relative to the systemic velocity by
100\,km/s, indicating the presence of some outflowing gas.
Alternatively, these two troughs, symmetric around the 6035\,MHz line,
might suggest the presence of gas rotating around a central source.
The VLBI jets in this source, a Seyfert~2 galaxy 16\,Mpc away, extend
to 1.5\,pc from the central engine \cite{14}; the 1.6\,GHz absorption
in this source has been found to come from nuclear region \cite{15}
and in addition to the broad velocity width, suggest that the 6\,GHz
absorption is also likely to come from the central region, where the
torus is expected to be. No absorption is observed in the 4.7\,GHz
transition.  \\ The absorption spectrum in NGC~5793 also shows a very
broad line width, which ranges up to 1000\,km/s (Fig.1, right
panel). Two narrow troughs are observed corresponding to the two
transitions at the systemic velocity and corresponding to a maximum
line opacity of $\tau\,\sim$\,0.034.\\ Both these sources show
previously detected OH in absorption in the lower state transitions at
1.6\,GHz
\cite{16}, \cite{17}.

\subsection{Cygnus A}

The 2\,cm continuum image of Cygnus~A (Fig.~2) shows a compact radio
source, extending east-west up to 4\,pc from the core, with a peak
flux density of 453\,mJy/beam. The noise in the map is
5\,mJy/beam. Fig.~2 shows the spectra for different positions at the
source. We report a tentative detection of the rotationally excited OH
transition at 13.434\,GHz towards the centre of Cygnus~A. The apparent
optical depth derived from the ratios of intensities of peak
absorption and adjacent continuum is $\tau$\,=\,0.125 with a line FWHM
corresponding to $\sim$ 90\,km/s. The absorption profile towards the
lobes is suggestive that part of the gas is diffuse and is surrounding
the inner jets, whereas a deeper and broader absorption profile is
seen towards the core. The profile is strongest when integrated over
the entire area containing continuum emission, and again seems to
indicate that the gas is spread over the whole source, with prevalence
towards the central region. Further tests are needed to confirm this
result and to rule out a spurious feature due to low-level
instrumental effects only detectable after long integration times.

\section{Conclusions}
We detect rotationally excited, broad OH lines in absorption towards
five of the 27 sources observed with the 100m telescope in
Effelsberg. This yields a detection rate of 19\,\%, which is higher
than the detection rates achieved in previous surveys in the
literature. These previous studies mainly targeted red quasars, where
an infrared excess is indicative of large columns of dust towards the
line of sight (e.g. \cite{18}, \cite{19}). In our study, source
selection was for the first time based on X-ray column densities. The
observed line widths range from a few 100\,km/s to 2000\,km/s,
suggesting that the gas in all sources is close to the central region,
either rotating around the central engine, or infalling/outflowing. We
find that the new 6\,GHz detections have absorption in the ground
state transition at 1.6\,GHz. This does not support the hypothesis of
radiative excitation models alone to explain the previous lack of
molecular detections.  However, given the high x-ray absorbing columns
in our systems, the non-detections still need to be explained. Since
the nature of the absorbing material is unknown, one possibility is
that it may be non-molecular in most galaxies. \\ We find that in the
five systems with detections the lines are strong and were visible
after short integration times, whereas in the non-detections no lines
were visible even after longer integration times. This bimodal
distribution of absorptions could be explained in the case of compact
clouds crossing the line of sight in a few of the sources.  \\ We also
report a tentative OH detection at 13.4\,GHz towards the powerful,
nearby galaxy Cygnus~A. This result needs further study to be
confirmed. The compactness of the radio source at this frequency
indicates that the absorbing material is within 4\,pc along the line
of sight to the central engine, consistent with most recent results
based on high resolution IR observations, showing torus sizes which
are no more than a few parsecs, e.g. \cite{20}, \cite{21}.

\end{document}